\documentclass[a4paper,11pt]{article}
\usepackage{jheppub} % for details on the use of the package, please
\usepackage[T1]{fontenc} % if needed

\usepackage{slashed}
\usepackage{stmaryrd}
\usepackage{color}
\usepackage{mathrsfs}
\usepackage{yfonts}
\usepackage{rotating}
\usepackage{graphicx}
\usepackage{amsfonts}
\usepackage{amsmath}
\usepackage{mathrsfs}
\usepackage[american]{babel}
\usepackage{wasysym}
\usepackage{slashed}
\usepackage{pgfplots}
\usepackage{braket}
\usepackage{graphicx}
\usepackage{subfigure}

\newcommand{\g}[0]{\gamma}

\newcommand{\de}[0]{\delta}

\newcommand{\la}[0]{\lambda}

\newcommand{\ea}[1]{\begin{align}#1\end{align}}
\newcommand{\eq}[1]{\begin{equation}#1\end{equation}}

\newcommand{\ma}[1]{\mathcal{#1}}

\begin{document}
\title{Spontaneously broken symmetry restoration of quantum fields
in the vicinity of neutral and electrically charged black holes
}
\author[a]{Gon\c{c}alo M. Quinta}
\emailAdd{goncalo.quinta@ist.utl.pt}
\affiliation[a]{Centro de Astrof\'{\i}sica e Gravita\c c\~ao - CENTRA,
Departamento de F\'{\i}sica, Instituto Superior T\'ecnico - IST,
Universidade de Lisboa - UL, Avenida Rovisco Pais 1,
1049-001 Lisboa, Portugal}
\author[b]{Antonino Flachi}
\emailAdd{flachi@phys-h.keio.ac.jp}
\affiliation[b]{Department of Physics and Research and Education Center
for Natural Sciences, Keio University, Hiyoshi 4-1-1, Yokohama,
Kanagawa 223-8521, Japan}
\author[c]{Jos\'{e} P. S. Lemos}
\emailAdd{joselemos@ist.utl.pt}
\affiliation[c]{Centro de Astrof\'{\i}sica e Gravita\c c\~ao - CENTRA,
Departamento de F\'{\i}sica, Instituto Superior T\'ecnico - IST,
Universidade de Lisboa - UL, Avenida Rovisco Pais 1,
1049-001 Lisboa, Portugal.}

\abstract{
We consider the restoration of a spontaneously broken symmetry of an
interacting quantum scalar field around neutral, i.e., Schwarzschild,
and electrically charged, i.e., Reissner-Nordstr\"om, black holes in
four dimensions. This is done through a semiclassical self-consistent
procedure, by solving the system of non-linear coupled equations
describing the dynamics of the background field and the vacuum
polarization. The black hole at its own horizon generates
an indefinitely
high temperature which decreases to the Hawking temperature at
infinity.  Due to the high temperature in its vicinity,
there forms a bubble around the black hole in which the scalar field
can only assume a value equal to zero, a minimum of energy. Thus, in
this region the symmetry of the energy and the field is preserved. At
the bubble radius, there is a phase transition in the value of the
scalar field due to a spontaneous symmetry breaking mechanism.
Indeed, outside the bubble radius the temperature is low enough such
that the scalar field settles with a nonzero value in a new energy
minimum, indicating a breaking of the symmetry in this outer
region. Conversely, there is symmetry restoration from the outer
region to the inner bubble close to the horizon.  Specific
properties that emerge from different black hole electric charges are
also noteworthy.  It is found that colder black holes, i.e., more
charged ones, have a smaller bubble length of restored symmetry.  In the
extremal case the bubble has zero length, i.e., there is no
bubble.  Additionally, for colder
black holes, it becomes harder to excite the quantum field modes, so
the vacuum polarization has smaller values. In the extremal case, the
black hole temperature is zero and the vacuum polarization is never
excited.
}

\maketitle

\section{Introduction}
\label{Int}

Early results on the stability of the universe and its possible
vacuum
decay through symmetry
breaking~\cite{Kobzarev:1974cp,Coleman:1977py,Callan:1977pt,Coleman:1980aw}
showed that our false vacuum could break and transit into a different
true vacuum.  With the discovery of the Higgs particle at the LHC
\cite{Higgs1,Higgs2} many new questions regarding the stability of our
universe have been brought up
\cite{Degrassi:2012ry,Espinosa:2015qea,Branchina:2014rva}.

Aside the universe, black holes may also trigger vacuum decay and act
as gravitational impurities able to nucleate in their surroundings a
true vacuum phase encased in a phase of false vacuum.  Thus, these
black holes act as nucleation sites causing vacuum restoration through
an inverted symmetry breaking process. These processes are
contemplated in scenarios of sufficiently hot black holes, where a
symmetric high temperature phase of a scalar field $\hat\phi$ say,
like the Higgs field, that forms in the vicinity of the evaporating
black hole supports the formation of a bubble of high temperature
phase.  For scalar fields with $\lambda \hat\phi^4$ interactions,
where $\lambda$ is some coupling, in Schwarzschild black holes
backgrounds, the problem was discussed in \cite{Hawking:1980ng}, where
it was argued that symmetry restoration of a broken phase at infinity
is expected to take place near a black hole horizon. However, the
initial conclusion was that in the Higgs model the region of symmetric
phase would be too localized for symmetry to practically be
restored. This problem was further examined
in~\cite{Fawcett:1981fw,Moss:1984zf,Hiscock:1987hn} where more
detailed calculations were carried out and the problem addressed to
different degrees, with the conclusion that sizable bubbles may
indeed form. Numerical lattice quantum Monte Carlo methods have been
applied to this problem in~\cite{Benic:2016kdk} for the a $\lambda
\hat\phi^4$ theory on a Schwarzschild black hole background leading to
a seemingly phase of broken symmetry in the near-horizon region.
Similar analyses in the framework of QCD phase transitions and chiral
symmetry breaking has been considered
in~\cite{Flachi:2011sx1,Flachi:2011sx2,Flachi:2015fna}.  Understanding
the birth and fate of black hole bubbles, given the right conditions
and the right kind of black holes, has always been a question of great
interest, much more now that it has been found out that the
probability that our false vacuum universe could transit into a
different true vacuum might be relatively high due to enhanced
nucleation from black hole
seeds~\cite{Gregory:2013hja,Burda:2015yfa1,Burda:2015yfa2}, when
compared with the predictions of the initial
works~\cite{Kobzarev:1974cp,Coleman:1977py,Callan:1977pt,Coleman:1980aw}.

The physics of the breaking, or of the restoration, is in fact
remarkably simple. Due to gravitational redshift, the radiation
emitted by a black hole looses energy and its temperature decreases as
it propagates through spacetime to distances far away from the
horizon. Inversely, from infinity to the horizon the temperature is
blueshifted and so increases. This makes it possible for a system in a
broken phase at large distances to have its symmetry restored
sufficiently close to the black hole. The local temperature becoming
larger than the characteristic critical temperature of the phase
transition, provides then a rationale for understanding in what
situations symmetry may be locally broken or restored.  Although
sufficient for this problem, this is not the entire story,
see~\cite{Flachi:2014jra,Castro:2018iqt}.

In this work, we examine this problem and construct bubble solutions
for $\lambda \hat\phi^4$ interactions and electrically charged,
i.e., Reissner-Nordstr\"om, black holes
adopting a semi-classically self-consistent approach that we implement
numerically.  These bubble solutions are solitons with the scalar
field changing abruptly from zero near the black hole horizon to some
finite value at some definite bubble radius.  The setup we consider
generalizes previous
results~\cite{Hawking:1980ng,Fawcett:1981fw,Moss:1984zf,Hiscock:1987hn}
in that we allow for the presence of a charge and our numerical
approach differs from that of~\cite{Benic:2016kdk}.  We use techniques
and results of
\cite{Parker:2009uva,Candelas:1980zt,Anderson:1989vg,Taylor:2017sux,Hewitt}.
A technical bonus of this work is that we chose to compute the quantum
vacuum polarization using the approach developed
in~\cite{Taylor:2017sux}, therefore putting to test in the present
context this novel computational method. This requires some
generalizations.

The paper is organized as follows.  In Sec.~\ref{PhySet}, the physical
and mathematical settings are described, the two main equations, one for
the background field and the other for the vacuum expectation value of the
quantum scalar field, i.e., the vacuum polarization, are derived.
In Sec.~\ref{SymR0}, we specify a generic
static spherical symmetric black hole background spacetime, display
the background field and the vacuum polarization equations for this
case outlining the procedure to calculate a renormalized result, study
the boundary conditions and the possibility of symmetry restoration.
In Sec.~\ref{NumRes}, we present the results for Schwarzschild
and
Reissner-Nordstr\"om spacetimes
and comment on the peculiarities and interest
of the solutions found. In Sec.~\ref{Conc}, we draw our
conclusions.

\section{Physical and mathematical settings}
\label{PhySet}
The physical situation
we consider here is that of a massive self-interacting
quantum scalar field $\hat{\phi}$
in a four-dimensional curved spacetime
containing a black hole, governed by the action
operator $S[\hat{\phi}]$ given by
\eq{\label{Sphi}
S[\hat{\phi}] = -  \int dv_x \left\{{1\over 2}
g^{ab}\hat{\phi}_{,a}\hat{\phi}_{,b} + {1\over 2}(\mu^2+
\xi R) \hat{\phi}^2 - {\la
\over 4} \hat{\phi}^4 \right\}\,,
}
where $dv_x \equiv \sqrt{|g|}\, d^4x $ is the four-dimensional
invariant volume element in curved spacetime, $g_{ab}$ is the metric
and $g$ its determinant, $a,b$ are four-dimensional
spacetime indices, a comma means partial derivative, $\mu$ is the mass
of the scalar field, $\xi$ and $\lambda$ coupling
parameters, and $R$ is the
Ricci scalar built out of the metric $g_{ab}$.
All the fundamental constants are set to one.

We start by expressing the quantum field $\hat{\phi}$ as excitations
over the background classical field, i.e., we write
\eq{\label{redef}
\hat{\phi}(x) =  \Phi(x) + \hat{\varphi}(x)\,,
}
where $\Phi =
\braket{\hat{\phi}}$ is the background classical field
and $\hat{\varphi}$ is the excitation  quantum field. The background
field $\Phi$ gives us information about the symmetry breaking of the
scalar field around the black hole and, as such, it is the quantity we
wish to calculate in the end. The action operator
given in
Eq.~(\ref{Sphi}) under the new field redefinitions
of Eq.~(\ref{redef})  can be
expressed as
\eq{\label{sexp}
S[\hat{\phi}] = S[\Phi] + {1 \over 2}
S_{,\Phi \Phi}[\Phi]\,\hat{\varphi}^2 + S_{\rm int}[\Phi,\hat{\varphi}]\,,
}
where
$S[\Phi]$ is $S[\hat{\phi}]$ evaluated at $\hat{\phi}=\Phi$,
the
notation $(\ldots)_{,\Phi'}$ denotes a functional derivative with
respect to the field $\Phi(x')$ with $x'$ labeling the spacetime
points of the field, and
$S_{\rm int}[\Phi,\hat{\varphi}]$ contains all the terms which
are not of order zero or two in the field $\hat{\varphi}$.

The background classical field
$\Phi$ is by definition
the solution of the functional differential equation
\eq{\label{GaEq}
{\de \Gamma[\Phi(x)] \over \de \Phi(x')} = 0
}
where $\Gamma$ is the effective action calculated using
Eq.~(\ref{sexp}).
We shall be concerned only with the lowest order
correction to the effective action, i.e.,
with 1-loop corrections, which is
equivalent to neglecting the term $S_{\rm int}$
in Eq.~(\ref{sexp}). One may then obtain
\cite{Parker:2009uva}
\eq{\label{EA}
\Gamma[\Phi] = S[\Phi] + {i \over 2}
\ln\left(\ell^2 S_{,\Phi \Phi}[\Phi]\right)\,,
}
where $\ell$ is an arbitrary constant introduced to keep the logarithm
dimensionless.

Now we define the Green function $G(x,x'')$ by
\eq{\label{GFdef}
\int dv_{x'} S_{,\Phi \Phi'} G(x',x'') = -\de^4(x,x'')\,,
}
where $\de^4(x,x'') \equiv \de^4(x-x'')/\sqrt{|g|}$ is the Dirac
delta-function in curved space. We note that
by definition the vacuum polarization of
the scalar field  $\braket{\phi^2}$ is the expectation value
of the square of the field in the coincidence
limit $x'\to x$, i.e.,
\eq{\label{exvac}
\braket{\phi^2(x)}\equiv -i G(x,x)\,.
}
We may then insert Eq.~(\ref{EA}) into
Eq.~(\ref{GaEq}) and make use of Eqs.~(\ref{GFdef}) and
(\ref{exvac}), to obtain
\eq{\label{DifEqPhi}
\square \Phi - \Phi\left(\la\Phi^2 - (\mu^2+\xi R) +
3\la \braket{\phi^2}\right)=0\,,
}
where $\square$ is the d'Alembertian.
Eq.~(\ref{DifEqPhi})
is the equation we need to solve to find
the background field $\Phi$.
We should remark that the effective action Eq.~(\ref{EA}) is
renormalizable, so the field quantities that appear in
Eq.~(\ref{DifEqPhi}) are to be interpreted as the finite, i.e.,
renormalized,
ones. In the above expression only the vacuum polarization
$\braket{\phi^2}$ diverges
and needs to be regularized~\cite{Parker:2009uva}.

We also need to write Eq.~(\ref{GFdef}) in a differential
operator form. To do that we
need to choose the vacuum state. Here, we
shall make the simplifying assumption that the black hole is in
thermal equilibrium with its environment, corresponding to choosing
the vacuum to be the Hartle-Hawking one. A more accurate description
would need the use of the Unruh vacuum state~\cite{Candelas:1980zt};
however, the present approximation, as
remarked in~\cite{Moss:1984zf}, is good in the case of bubble walls
larger than the predominant wavelength of the radiation.
With this choice Eq.~(\ref{GFdef}) becomes
\eq{\label{GFeq}
(\square - \mu^2 - \xi R + 3\la \Phi^2)G(x,x') = - \de^4(x,x')\,.
}
Since the limiting value of the Green function, i.e., the vacuum
polarization $\braket{\phi^2}$, depends on the background field itself,
we will have a
system of non-linear coupled differential equations comprised of
Eqs.~(\ref{DifEqPhi}) and Eq.~(\ref{GFeq}). Physically, what we have
is essentially a vacuum polarization that must be calculated for an
effective mass squared $Q$, say, given by
$Q=\mu^2+\xi R-3\la \Phi^2$.
We will show  how to find a very
good approximation to the solution of
the two dynamical equations~(\ref{DifEqPhi}) and~(\ref{GFeq})
in the case of a static spherical symmetric
background.

To understand the possibility of a phase transition
or symmetry restoration
we write the effective action Eq.~(\ref{EA})
as
\eq{\label{Sphieffe}
\Gamma[{\Phi}] = -  \int dv_x \left\{
{1\over 2}g^{ab}{\Phi}_{,a}{\Phi}_{,b}- U\right\}\,,
}
where the  effective potential $U(\Phi)$ is defined as
\eq{\label{Ueff}
U= -{1\over 2}(\mu^2+\xi R-3\la \braket{\phi^2})\Phi^2
+ {1\over 4} \la \Phi^4\,.
}
First, note that the effective potential $U$ is symmetric
around $\Phi=0$.
Second, from its derivative,
\eq{\label{dUeff}
\frac{dU}{d\Phi}= -(\mu^2+\xi R-3\la \braket{\phi^2})\Phi
+  \la \Phi^3\,,
}
note
that if $\braket{\phi^2}$ is low, or even zero as in the
pure classical case, then $U$ has one local maximum
at $\Phi=0$ and one
global minimum at a nonzero value of
$\Phi$. Thus, the system sets
at this nonzero value of
$\Phi$ for which $U$ is a global minimum,
breaking the symmetry of $U$.
Third, note however, that if $\braket{\phi^2}$ is high enough
we see from  Equation~(\ref{dUeff})
that there is
only one global minimum for $U$ which occurs at
$\Phi=0$. Thus, the system sets
at this zero value of
$\Phi$ for which $U$ is a global minimum and the symmetry of
$U$ is restored from quantum processes.
A black hole at a given temperature in equilibrium
with a quantum field can realize this
symmetry restoration. Far way from the black hole
the temperature is low enough and the field
is at the
global minimum with a nonzero value
in a symmetric broken phase, near the
black hole the temperature is high enough,
$\braket{\phi^2}$ is important, and
the field changes its minimum to $\Phi=0$,
restoring the symmetry.

\section{Dynamical equations}
\label{SymR0}

\subsection{Background field equation}
\label{SymR}

We assume a scalar quantum field
$\hat \phi=\Phi+\hat\varphi$ sitting in
a static spherical symmetric
spacetime background.
We want
to study the thermal properties of the
quantum field and so we work
with a Euclidean time $t$.
Since spacetime is
spherically symmetric we choose
as coordinates
$t,r,\theta,\varphi$, with $r$
being the radial coordinate and
$\theta$ and $\phi$ the angular ones.
The  Euclidean line element is then written as
\eq{\label{Emetric}
ds_{\rm E}^2 = f(r)dt^2 + {1 \over f(r)} dr^2 +
r^2 \left( d\theta^2+\sin^2\theta d\varphi^2\right)\,,
}
where $f(r)$ is some function of $r$.
We further consider that $f(r)$ represents a black hole.
Since the spacetime is spherically symmetric, we shall make the
assumption that our regular background
field configuration is also
spherically symmetric, that is, depends solely on the radial
coordinate, i.e.,
\eq{\label{Escalarsph}
\Phi \equiv \Phi(r)\,.
}

For the Euclideanized metric Eq.~(\ref{Emetric})
and a field of the type given in Eq.~(\ref{Escalarsph}),
the background field will then be a solution of Eq.~(\ref{DifEqPhi}),
now in the form
\eq{\label{EDifEqPhi}
\bigg\{{d^2 \over dr^2} + \left({2 \over r}+{f' \over f}
\right){d \over dr}\bigg\}\Phi - \left({\mu^2+\xi R-3\la \braket{\phi^2}
\over f}\right)\Phi + {\la \over f}\Phi^3=0\,.
}
The differential equation (\ref{EDifEqPhi}) has some interesting
features which make it fairly challenging to directly obtain a numeric
solution \cite{Hawking:1980ng}.
Moreover, Equation~(\ref{EDifEqPhi})
has to be solved consistently with the
vacuum polarization equation.

%%%%%%%%%%%%%%%%%%%%%%%%%%%%%%%%%%%%%%%%%%%%%%%%%%%%%%%%%%%%%%%%%%%%%%%%%%%%%%
\subsection{Vacuum polarization equation}
\label{VacP}

Performing
a Wick rotation on Eq.~(\ref{GFeq}), we obtain
\eq{\label{EGFeq}
(\square_{\rm E} - Q(r))G_{E}(x,x') = - \de^4(x,x')\,,
}
where
$\square_{\rm E}$ is the Euclidean d'Alembertian, i.e.,
the Laplacian operator of the Euclidean space
with metric Eq.~(\ref{Emetric}),
\eq{\label{EGFeq0}
G_{\rm E}(x,x') = i G(x,x')\,,
}
is the appropriate Euclidean Green function,
and
$Q$ is taken to be a generic radial dependent mass squared term.
In our case, from Eq.~(\ref{GFeq}), we get
\eq{\label{Q}
Q(r)=\mu^2 + \xi R - 3\la \Phi^2\,,
}
so that $Q$ is the effective mass
squared.
From Eq.~(\ref{exvac}), the vacuum polarization will now be given by
\eq{\label{phig}
\braket{\phi^2(x)} =  G_{\rm E}(x,x)\,.
}

The solution of Eq.~(\ref{EGFeq}) for a field in thermal equilibrium
with a black hole can be decomposed into energy modes $n$ and angular
modes $l$  in the form \cite{Anderson:1989vg}
\eq{
G_{\rm E}(x,x') = {T_{\rm H}\over 2\pi} \sum^{\infty}_{n=-\infty}
\left(l+{1\over 2}\right) P_l(\cos\g) g_{nl}(r,r')\,,
}
where $T_{\rm H}$ is the Hawking temperature,
i.e., the black hole temperature at
infinity, $P_l(x)$ are the
Legendre polynomials, $\g$ is the geodesic distance on the 2-sphere
defined by $(\theta,\varphi)$
and $g_{nl}$ are the radial Green function modes, which
from Eq.~(\ref{EGFeq})
satisfy
\eq{\label{RGF}
\bigg[{d^2 \over dr^2} + \left({2 \over r}+
{f' \over f}\right){d \over dr} - \bigg({l(l+1) \over r^2 f}
 + {4\pi^2 T_{\rm H}^2 n^2 \over f^2} + {Q \over f}
\bigg)\bigg] g_{nl}(r,r') = - \de(r-r')\,.
}
The mode functions $g_{nl}$ can be expressed as
\eq{
g_{nl}(r,r') = N_{nl} \, p_{nl}(r_<) q_{nl}(r_>)\,,
}
where $p_{nl}$ and $q_{nl}$ are the homogeneous solutions of
Eq.~(\ref{RGF}) regular at the horizon and infinity, respectively. The
$N_{nl}$ is a normalization constant, given by
\eq{
N_{nl} = -{1\over r^2 f(r) \mathcal{W}(p_{nl},q_{nl})}\,,
}
where $\mathcal{W}(p_{nl},q_{nl})$ is the Wronskian of the two solutions.
We also use the notation $r_{<} \equiv \textrm{min}\{r,r'\}$ and $r_{>}
\equiv \textrm{max}\{r,r'\}$.

The mode functions $g_{nl}$ are divergent in the coincidence limit,
and hence so is the vacuum polarization. In order to find a
physically meaningful result, one must apply a renormalization
procedure to obtain a finite quantity. In this work we will employ the
method developed in \cite{Taylor:2017sux}. The process is quickly
convergent at the horizon, so it is appropriate to our situation where
multiple instances of $\braket{\phi^2}$ will have to be
calculated. The procedure essentially applies a very specific choice
of point-splitting which allows the isolation of the divergent, or singular,
piece
of the mode functions, denoted as $g^S_{nl}$, which is then subtracted
from a numerical calculation of $g_{nl}$ to give a finite result.
In
the end, the renormalized vacuum polarization
$\braket{\phi^2(r)}_{\textrm{ren}}$ becomes written as
\eq{\label{RenVac}
\braket{\phi^2(r)}_{\textrm{ren}} = {T_{\rm H}\over 4\pi}
\sum^{\infty}_{l=0} (2l+1) \bigg\{g_{0l}(r)-g^{S}_{0l}(r) + 2\sum^{\infty}_{n=1}\left(g_{nl}(r)-
g^S_{nl}(r)\right)\bigg\} + {f'(r) \over 48\pi^2}
}
with
\eq{
g^S_{nl}(r) = {2\pi\over T_{\rm H}}\bigg\{\sum^{2}_{i=0}\sum^{i}_{j=0}
\mathcal{D}^{(+)}_{ij}(r) \Psi^{(+)}_{nl}(i,j|r) +\mathcal{T}^{(r)}_{10}(r) \Psi^{(-)}_{nl}(2,0|r)+\sum^{1}_{i=0}
\sum^{i}_{j=0} \mathcal{T}^{(l)}_{ij}(r)\chi_{nl}(i,j|r)\,.
}
We have derived the quantities $\mathcal{D}^{(+)}_{ij}(r)$,
$\mathcal{T}^{(r)}_{10}(r)$ and $\mathcal{T}^{(l)}_{ij}(r)$ for a
general mass term squared, $Q(r)$, which are listed in the Appendix A.
In our specific case $Q(r)$ is given in Eq.~(\ref{Q}).
The
functions $\Psi^{(+)}_{nl}(i,j|r)$ and $\chi_{nl}(i,j|r)$ are rather
lengthy, and are exactly the same as in \cite{Taylor:2017sux}. For
details on the form and derivation of these results, we refer the
reader to consult \cite{Taylor:2017sux}.

\subsection{Boundary conditions and symmetry restoration}
\label{bcSymR2}

We assume that the
function $f(r)$ in Eq.~(\ref{Emetric}) yields a black hole with a
horizon, the spacetime
is asymptotically flat, and has a Ricci scalar $R=0$.
We also assume that the black hole temperature at infinity
is the Hawking temperature
$T_{\rm H}$.

\begin{figure}
\centering
\includegraphics[width=0.4\textwidth]{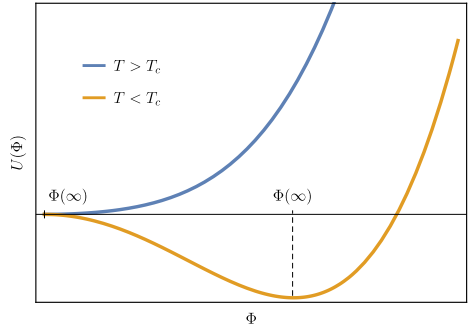}
\caption{Profile of the effective potential $U(\Phi)$ at $r=\infty$,
for $T>T_c$ and $T<T_c$, as a function of $\Phi$. The minimum of the effective potential corresponds to $\Phi(\infty)$. For $T>T_c$, the only possible minimum is the trivial one
$\Phi(\infty)=0$.}
\label{Boundary}
\end{figure}

We have to impose a boundary condition at infinity.  At very large
radii, we have that for asymptotically flat spaces at any temperature
$T$ and any field mass $\mu$ the vacuum polarization
$\braket{\phi^2}$ is given by
$
\braket{\phi^2(\infty)} = {1\over 2\pi^2} \int^{\infty}_{m}
{\sqrt{\omega^2-\mu^2} \over e^{\omega / T} - 1} \, d\omega
$ \cite{Hewitt}.
In our case, at infinity the temperature is the
Hawking temperature, so we put $T=T_{\rm H}$, yielding,
\eq{
\braket{\phi^2(\infty)} = {1\over 2\pi^2} \int^{\infty}_{m}
{\sqrt{\omega^2-\mu^2} \over e^{\omega / T_{\rm H}} - 1} \, d\omega\,.
}
Note that $\braket{\phi^2(\infty)} $ is a function of $T_{\rm H}$
and $m$.
Also, at large radii, the derivative terms in
Eq.~(\ref{EDifEqPhi}) are negligible, so we can solve for the field
$\Phi$, obtaining after recalling that we put $R=0$,
\eq{\label{PhiInf}
\Phi(\infty) = \sqrt{{\mu^2 \over \la} - 3 \braket{\phi^2(\infty)}}
\quad \textrm{for} \quad T_{\rm H} \leq T_c\,,
}
where $T_c$ is a critical temperature defined at
infinity as the
temperature for which the square root of Eq.~(\ref{PhiInf}) becomes
negative.
For $T_{\rm H}  <T_c$ one has that
$\mu^2 - 3\la \braket{\phi^2(\infty)}>0$ so
$\Phi(\infty)$ is positive. In addition,
from
Eq.~(\ref{Ueff}),
the effective potential $U$
has a local maximum and a global minimum,
see Fig.~\ref{Boundary}.
Since the temperature at any other radius
is blueshifted from $T_{\rm H}$ up to
and infinite temperature at the horizon,
it is possible that
the temperature at a certain radius $r$
will be sufficiently high so that
the effective potential $U$ will
have
global minimum only, signaling a phase transition,
i.e., symmetry restoration.
On the other hand,
for $T_{\rm H}  >T_c$,  one has that
$\mu^2 - 3\la \braket{\phi^2(\infty)}<0$, and
Eq.~(\ref{PhiInf}) gives that the
field $\Phi$ is imaginary, so in fact one should pick
the trivial solution of
Eq.~(\ref{EDifEqPhi}) at infinity, i.e.,
\eq{
\Phi(\infty) = 0  \quad \textrm{for} \quad T_{\rm H} >T_c\,.
}
Thus, for such high temperatures at infinity,
from
Eq.~(\ref{Ueff}),
the effective potential $U$ there  has a
global minimum only at $\Phi(\infty)=0$.
Since the temperature at any other radius
is blueshifted from $T_{\rm H}$,
the temperatures in the whole region up
to the horizon are always higher than
the Hawking temperature at infinity so
in principle there
is no qualitative change in the
effective potential $U$ and there is no
possibility of symmetry breaking.
This case is in this sense trivial and
we are not interested
in  it. We want symmetry restoration at some point
$r$ from
infinity to the horizon. So we deal with
$T_{\rm H}  <T_c$ and
Eq.~(\ref{PhiInf}) is the one that interests us.
The result
Eq.~(\ref{PhiInf}) will thus be used as the boundary condition at
infinity.

\begin{figure}
\centering
\includegraphics[width=0.4\textwidth]{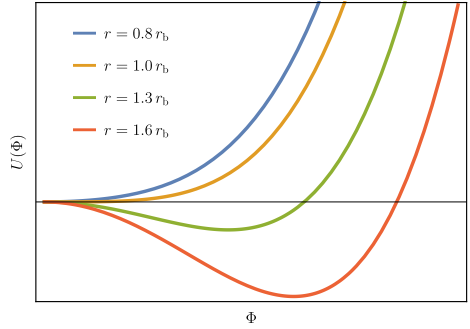}
\caption{Sketch of the profile of the effective potential $U(\Phi)$ for decreasing
radii from $r=\infty$. The minimum of the potential is negative for $r>r_{\rm
b}$. From
$r<r_{\rm b}$ and inwards, the minimum of the potential is zero,
so the background field is $\Phi=0$ in that range.
}
\label{bubble}
\end{figure}

In order to better understand the symmetry restoration
we recall the effective potential
$U(\Phi)$ that appears naturally in
Eq.~(\ref{Ueff}).
We sketch in Fig.~\ref{bubble}
the plots of
$U(\Phi)$
as a function of $\Phi$ for
several different radii $r$ in the case that $T_{\rm H}$
is less
than $T_c$ at infinity.
For a large radius $U(\Phi)$
as a function of $\Phi$ shows the same behavior that as
at infinity, i.e.,
it has a local maximum and a global minimum.
But, at a certain
radius we have that $\braket{\phi^2}$ achieves
a value
such that
the global minimum of $U(\Phi)$ in Eq.~(\ref{Ueff}) is at
$\Phi=0$, and so symmetry is
restored, see Fig.~\ref{bubble}. This radius
we call
the bubble radius $r_{\rm b}$.

\section{Numerical solutions for Schwarzschild and
Reissner-Nordstr\"om black holes}
\label{NumRes}

\subsection{Results, plots, and analysis}

In this work, we will be concerned with the spontaneously broken
symmetry restoration of a scalar field around a non-charged and a
charged four dimensional black hole, whose geometries are described by
the Schwarzschild and Reissner-Nordstr\"om metrics, respectively,
namely, $f(r)$ in Eq.~(\ref{Emetric}) takes the form
\eq{
f(r) =
\left(1-\frac{r_+}{r}\right)
\left(1-\frac{r_-}{r}\right)\,,
}
where $r_+$ is the horizon radius and $r_-$ the Cauchy radius.
In terms of the ADM mass $m$ and electrical charge $q$
these are given by
$r_{+} = m + \sqrt{m^2-q^2}$ and
$r_{-} = m - \sqrt{m^2-q^2}$.
For $q=0$ one $r_+=2m$, $r_-=0$ and
the Schwarzschild space is recovered,
$f(r)=1-\frac{r_+}{r}$.
For $q=m$ one gets $r_+=r_-=m=q$, and
the the extremal Reissner-Nordstr\"om space is obtained,
$f(r)=(1-\frac{r_+}{r})^2$.
We assume that $r_+\geq r_-$, i.e.,
$m\geq q$, so that there
is always an $r_+$ and thus there are no naked singularities.
Since for the Reissner-Nordstr\"om metric
the space geometry satisfies $R=0$,
the curvature coupling $\xi R$ is irrelevant.
The black hole temperature is the Hawking temperature
$T_{\rm H}$ given for the  Reissner-Nordstr\"om space by
\eq{\label{BHtemp}
T_{\rm H} = {1 \over 4\pi r_+}\left(1-\frac{r_-}{r_+}\right)\,.
}
Schwarzschild black holes, $r_-=0$, are the hottest.
Extremal black holes, $r_-=r_+$, are the coldest, have zero temperature.
Indeed, as the Cauchy
radius $r_-$ is increase, i.e., as the black hole electric
charge increases,
the black hole becomes colder, see
Eq.~(\ref{BHtemp}).

The system of equations (\ref{EDifEqPhi}) and (\ref{EGFeq}) has not
an exact solution, so we must resort to an approximation scheme. To
solve this system, we will employ an approximation that is self-consistent
semi-classically. The procedure is as follows. First,
from Eq.~(\ref{RenVac}), which is a development on Eq.~(\ref{EGFeq})
with Eq.~(\ref{phig}) inserted,
we compute $\braket{\phi^2}_{\rm ren}$ for the case where no
background field is present, i.e., $\Phi=0$ and so
$Q(r) = \mu^2$, from here onwards we put
$\xi R=0$. Second, we insert
the result into Eq.~(\ref{EDifEqPhi}) and compute the resulting
$\Phi$.
This step is nontrivial and will be
detailed below.
Third, we compute again the vacuum polarization but now
with the new $\Phi$ included in Eq.~(\ref{RenVac}), i.e., such that
$Q(r) = \mu^2- 3 \la
\Phi^2(r)$. Fourth,
we take the resulting $\braket{\phi^2}_{\rm ren}$ and put
it back into Eq.~(\ref{EDifEqPhi}), giving a new function for
$\Phi$. These steps are repeated until the results for the
background field $\Phi$
and
the vacuum
polarization $\braket{\phi^2}_{\rm ren}$
stop changing appreciably.
This is the self-consistent procedure.

We now analyze in detail the numerical procedure for solving
Eq.~(\ref{EDifEqPhi}).  In order to solve Eq.~(\ref{EDifEqPhi})
numerically, for each iteration of the self-consistent approximation,
we divide the problem into three stages: First we find the bubble
radius approximately, second we solve Eq.~(\ref{EDifEqPhi})
from the horizon to the
bubble radius, third we solve Eq.~(\ref{EDifEqPhi})
from the bubble radius to infinity. The
three stages spelled out are as follows.  First, to find the bubble
radius, for the given $\braket{\phi^2}_{\rm ren}$, we search from the
inside by trial
and error for a radius for which for the first time the effective
potential shows a minimum at some nonzero $\Phi$.  This gives an
approximate bubble radius $r_{\rm b\,approx}$ and a field $\Phi$ at
the approximate bubble radius $\Phi(r_{\rm b\,approx})$.
Second, to solve
Eq.~(\ref{EDifEqPhi}) from the horizon to the bubble radius, i.e.,
inside the bubble, we choose a point very close to the horizon,
essential $r_+$, and evaluate the minimum of the effective potential
for that radius, obtaining $\Phi(r_+)$. Using this value $\Phi(r_+)$
together with the determined value of the background field at the
approximate bubble radius $\Phi(r_{\rm b\,approx})$, we find the
solution inside the bubble using Eq.~(\ref{EDifEqPhi}).  Third, to
solve from the bubble radius to infinity, i.e., for the region outside
the bubble, it remains to find the value of the field at infinity,
i.e., for some sufficient large value of the radius. Since the field
will be considered to be at thermal equilibrium with the black hole,
its value at infinity $\Phi(\infty)$ will be given by
Eq.~(\ref{PhiInf}) at the black hole temperature, i.e., $T_{\rm H}$,
with $T_{\rm H}$ given by Eq.~(\ref{BHtemp}).  Using this value of
$\Phi(\infty)$ together with the determined value of the background
field at the approximate bubble radius $\Phi(r_{\rm b\,approx})$, we
find using Eq.~(\ref{EDifEqPhi}) the solution outside the bubble.
Thus, for a given $\braket{\phi^2}_{\rm ren}$ we find $\Phi(r)$ for
the whole space.  We then resort to the next step in the
self-consistent approximation until the solution stops changing
appreciably.

\begin{figure}[h]
\centering
\includegraphics[width=0.5\textwidth]{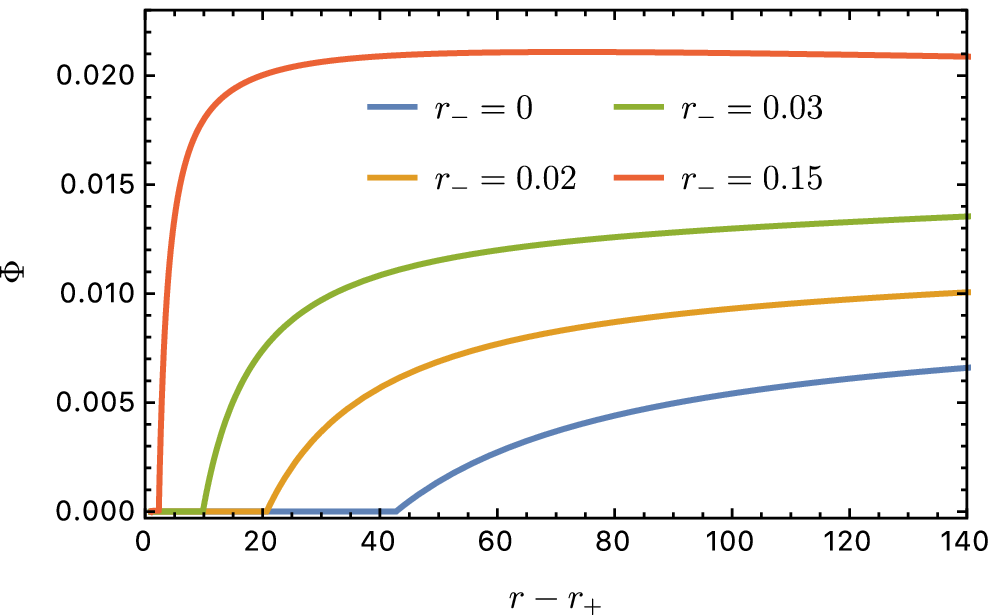}
\vskip -0.3cm
\caption{Profile of the background field $\Phi$ for $\la = 7.1\times 10^{-3}$,
$\mu = 0.01$, $r_+=1$ and varying $r_-$. The case $r_-=0$ is
the Schwarzschild case and  $r_-=0.02,0.03,0.15$ are
typical Reissner-Nordstr\"om cases. The extremal case $r_-=1$,
not shown, starts from $\Phi=0$ at $r-r_+=0$ and jumps immediately
to a finite value in a step function as can be inferred
from the plots, i.e., there is no bubble for extremal black holes.}
\label{1}
%\end{figure}
%\begin{figure}[h]
\vskip 0.8cm
\centering
\includegraphics[width=0.5\textwidth]{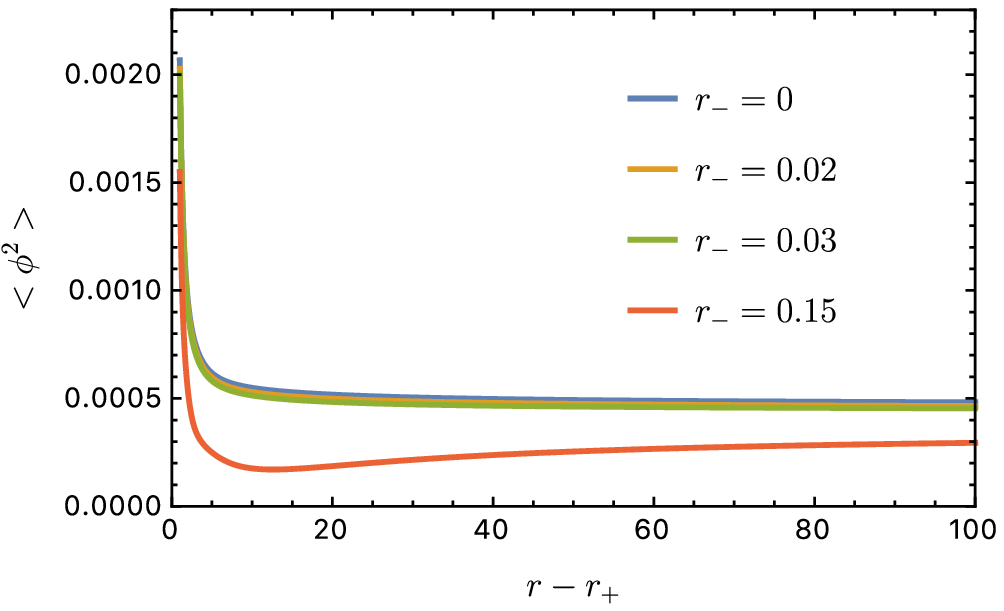}
\vskip -0.4cm
\caption{Profile of the vacuum polarization $\braket{\phi^2}_{\rm ren}$
for $\la = 7.1\times
10^{-3}$, $\mu = 0.01$, $r_+=1$ and varying $r_-$. The case $r_-=0$ is
the Schwarzschild case and  $r_-=0.02,0.03,0.15$ are
typical Reissner-Nordstr\"om cases. The extremal case $r_-=1$,
not shown, gives  $\braket{\phi^2}_{\rm ren}=0$ for all $r-r_+$
as can be inferred
from the plots, i.e., there is no vacuum polarization
for extremal black holes.
}
\label{2}
\end{figure}

At this stage we have a well-defined bubble radius $r_{\rm b}$ and the
solution $\Phi(r)$ and $\braket{\phi^2}_{\rm ren}(r)$ throughout all
space.  Employing thus the self-consistent approximation we obtain the
results for the background field $\Phi(r)$ in Fig.~\ref{1} and for the
vacuum polarization $\braket{\phi^2}_{\rm ren}(r)$ in Fig.~\ref{2}. A
careful analysis of the results and plots is now in order.

From the point of view of the background field $\Phi$, spontaneously
broken symmetry restoration is less likely for colder black holes, so
colder black holes, i.e., more charged ones, will have a smaller
bubble of restored symmetry, something that is clear from
Fig.~\ref{1}.  For extremal black holes $r_-=r_+$ the bubble has zero
width, thus for extreme black holes there is no bubble, and the curve
is a step function.  Although not shown it is clear that this is the
limit of the curves drawn in Fig.~\ref{1}.

From the point of view of the vacuum polarization
$\braket{\phi^2}_{\rm ren}$, there are several aspects that one can
rise:
(i) For colder black holes it becomes harder to excite the quantum
field modes, so the vacuum polarization for those black holes has
smaller values as can be checked in Fig.~\ref{2} when comparing each
different curve $r_-$.  Extremal black holes $r_-=r_+$ have zero
temperature, the vacuum polarization is never excited, and so the
curve is the line $\braket{\phi^2}_{\rm ren}(r)=0$.  Although not
shown it is clear that this is the limit of the curves drawn in
Fig.~\ref{2}.
(ii) For each black hole, i.e., for a given $r_-$, we clearly see from
Fig.~\ref{2} that as $r$ increases, and thus the temperature
decreases, it also becomes harder to excite the quantum field modes,
so the vacuum polarization is smaller as $r$ increases.
(iii) The influence of the background field $\Phi$ for a given curve
$r_-$ at the horizon and at infinity is also worth analyzing.  At the
horizon, the background field $\Phi$ is negligible, so the value of
$\braket{\phi(r_+)^2}$ is unaltered by $\Phi$, the changes in
$\braket{\phi(r_+)^2}$ there come from other sources.  At infinity,
the field $\Phi$ has its largest value which translates into a smaller
effective mass squared $Q=\mu^2-3\la \Phi^2$ and in turn this
increases the vacuum polarization.
(iv) Another interesting fact is that the overall form of the vacuum
polarization is more affected by smaller bubbles.  This is because
$\braket{\phi^2}$ stabilizes quickly in a distance relatively small
from the horizon, so the background field can only alter the form of
the vacuum polarization curve when its region of larger variations,
i.e., the outer edge of the bubble, is situated near the horizon.  For
larger bubbles, we see from Fig.~\ref{2} that the effects of the
background field on the vacuum polarization are not so distinct.

\subsection{Comments on the numerical calculations}

Commenting on the numerical calculations, we observe that the
solutions stabilize relatively fast, at the order of three or four
iterations of the self-consistent approximation.

Regarding the vacuum polarization, the method employed here converges
quickly on the horizon, where only some tens of modes are necessary to
obtain a good result. However, at large radii, at the order of some
hundreds of $r_+$, the convergence for the vacuum polarization becomes
slower, requiring the sum of some hundreds of modes. Since the order
of magnitude of each Green mode function becomes very small for large
distances, we are faced with the task of calculating very accurately
hundreds of differences between very small numbers in
Eq.~(\ref{RenVac}). As a consequence, we must find the numerical
solutions of the homogeneous version of Eq.~(\ref{RGF}) for each mode
with a very high precision, which revealed to be a considerable heavy
and fine-tuned task for the symbolic manipulation software Mathematica
used by us for the purpose. These shortcomings increased the overall
computational time, which was reasonably lessened by parallelizing the
code and using it in a cluster. In this regard, the numeric efficiency
may be improved by adopting a different method to find the numerical
solutions for the mode functions.

The results have been further checked using a slightly different
approach which fixes the value of the field asymptotically and uses
the value of the derivative as a shooting parameter. We have verified
in a number of cases that the solutions obtained in the two ways
coincide to the numerical accuracy we have used.

%%%%%%%%%%%%%%%%%%%%%%%%%%%%%%%%%%
\section{Conclusions}
\label{Conc}

In this work we have constructed bubble solutions for a
self-interacting quantum scalar field around non-charged
and charged four
dimensional black holes. These bubble solutions can be envisaged as
solitons with $\Phi=0$ inside the bubble and $\Phi$ finite outside it.
The method we have adopted includes a self-consistent calculation of
one-loop quantum effects encoded in the scalar vacuum polarization.
The
results we have obtained clearly demonstrate the picture where a
spontaneously broken symmetry phase far away from the black hole
is restored sufficiently near its horizon
due to the increase of the local temperature associated to the
gravitational blueshift from the Hawking temperature at infinity to an
large unbound temperature at the horizon.  We have confirmed this picture
by extending the results of
\cite{Hawking:1980ng,Fawcett:1981fw,Moss:1984zf} and by explicitly
constructing the solutions for the bubble configuration. In
particular, we have observed that hot black holes
have large bubble regions where the
temperature is high enough to induce a phase transition in the value
of $\Phi$, cold black holes  have small bubbles, with
the extremal black hole, at zero temperature,
having no bubble, $\Phi$ is a nonzero constant throughout.

%\noindent
\section*{Acknowledgments}

We thank Peter Taylor and Cormac Breen for sharing the coefficients for a
constant mass term in four dimensions that helped us deriving
$\braket{\phi^2(x)}_{\textrm{ren}}$ and $g^S_{nl}(r)$ above.
GQ acknowledges the support of the Funda\c c\~ao para a Ci\^encia e
Tecnologia (FCT Portugal) through Grant No.~SFRH/BD/92583/2013.
AF acknowledges the support of the Japanese Ministry of Education,
Culture, Sports, Science Program for the Strategic Research Foundation
at Private Universities `Topological Science' Grant No.~S1511006 and
of the JSPS KAKENHI Grant No.~18K03626.
JPSL acknowledges FCT for financial support through
Project~No.~UID/FIS/00099/2013, Grant No.~SFRH/BSAB/128455/2017,
and Coordena\c c\~ao de Aperfei\c coamento do Pessoal de N\'\i vel
Superior (CAPES), Brazil, for support within the Programa CSF-PVE,
Grant No.~88887.068694/2014-00.
The authors thankfully acknowledge the computer resources, technical
expertise, and assistance provided by CENTRA/IST. Computations were
performed at the cluster Baltasar-Sete-Sois supported by the H2020 ERC
Consolidator Grant ``Matter and strong field gravity: New frontiers in
Einstein's theory" grant agreement No.~MaGRaTh-646597.

%\vfill\vskip 2cm
%%%%%%%%%%%%%%%%%%%%%%%%%%%%%%%%%%%%%%%%%%%%%%%%%%%%%%%%%%%%%%%%%%
\appendix
\section*{Appendix: Hadamard coefficients}
\label{SymB}
For a field with a radial dependent mass term squared,
$Q(r)$, we find the Hadamard coefficients
below. In the formulas, to avoid the appearance of
$2\pi$ and its powers an excessive number of times,
we use the surface gravity $\kappa$
of the black hole instead of its Hawking temperature
$T_{\rm H}$, the relation between them being
$T_{\rm H}={\kappa\over2\pi}$.
Denote differentiation with respect to the radial coordinate
with a prime. The Hadamard coefficients are
\ea{
\ma{D}^{(+)}_{00} = & \, 2\,, \\
\ma{D}^{(+)}_{10} = & \, \frac{f}{6 r^2}+\frac{f' f}{6 r}-\frac{1}{12} f'' f-\frac{f^2}{6 r^2} \,, \\
\ma{D}^{(+)}_{11} = & \, -\frac{1}{6} \kappa^2 f+\frac{1}{24} f'^2 f-\frac{f^2}{6 r^2}-\frac{f' f^2}{6 r}+\frac{f^3}{6 r^2} \,, \\
\ma{D}^{(+)}_{20} = & \, \frac{\kappa^2 f}{72 r^2}+\frac{\kappa^2 f' f}{72 r}-\frac{f'^2 f}{288 r^2}-\frac{f'^3 f}{288 r}-\frac{1}{144}
\kappa^2 f'' f+\frac{1}{576} f'^2 f'' f+\frac{f^2}{720 r^4}-\frac{\kappa^2 f^2}{72 r^2} \nonumber \\
& +\frac{f'
f^2}{36 r^3}+\frac{43 f'^2 f^2}{1440 r^2}-\frac{f'' f^2}{144 r^2} -\frac{7 f' f'' f^2}{360 r} +\frac{1}{320}
{f''}^2 f^2+\frac{1}{240} f' f^{(3)} f^2-\frac{f^3}{36 r^4} \nonumber \\
& -\frac{19 f' f^3}{360 r^3}+\frac{7 f''
f^3}{360 r^2}-\frac{f^{(3)} f^3}{120 r}+\frac{19 f^4}{720 r^4}  \,, \\
\ma{D}^{(+)}_{21} = & \, -\frac{1}{45} \kappa^4 f+\frac{1}{144} \kappa^2 f'^2 f-\frac{f'^4 f}{2880}-\frac{\kappa^2 f^2}{24 r^2}-\frac{\kappa
^2 f' f^2}{24 r}+\frac{f'^2 f^2}{96 r^2}+\frac{f'^3 f^2}{96 r}+\frac{1}{144} \kappa^2 f'' f^2 \nonumber \\
& -\frac{11
f'^2 f'' f^2}{2880}-\frac{7 f^3}{360 r^4}
+\frac{\kappa^2 f^3}{24 r^2}-\frac{f' f^3}{18 r^3}-\frac{67
f'^2 f^3}{1440 r^2}+\frac{f'' f^3}{144 r^2}+\frac{11 f' f'' f^3}{720 r}+\frac{f^4}{18 r^4} \nonumber \\
& +\frac{13
f' f^4}{180 r^3}-\frac{11 f'' f^4}{720 r^2}-\frac{13 f^5}{360 r^4}  \,, \\
\ma{D}^{(+)}_{22} = & \, \frac{1}{72} \kappa^4 f^2-\frac{1}{144} \kappa^2 f'^2 f^2+\frac{f'^4 f^2}{1152}+\frac{\kappa^2 f^3}{36 r^2}+\frac{\kappa
^2 f' f^3}{36 r}-\frac{f'^2 f^3}{144 r^2}-\frac{f'^3 f^3}{144 r}+\frac{f^4}{72 r^4}-\frac{\kappa^2
f^4}{36 r^2} \nonumber \\
& +\frac{f' f^4}{36 r^3}+\frac{f'^2 f^4}{48 r^2}-\frac{f^5}{36 r^4} -\frac{f' f^5}{36 r^3}+\frac{f^6}{72
r^4}  \,, \\
\ma{T}^{(l)}_{00} = & \, \frac{1}{24} Q \kappa^2 f-\frac{\kappa^2 f}{72 r^2}+\frac{\kappa^2 f' f}{36 r}-\frac{1}{96}
Q f'^2 f+\frac{f'^2 f}{288 r^2}-\frac{f'^3 f}{144r}+\frac{1}{144} \kappa^2 f^{(2)} f-\frac{1}{576} f'^2 f^{(2)} f \nonumber \\
& -\frac{f^2}{72r^4}+\frac{Q f^2}{24 r^2}+\frac{\kappa^2 f^2}{72 r^2} +\frac{f' f^2}{72
r^3}+\frac{Q f' f^2}{24 r}+\frac{7 f'^2 f^2}{288 r^2}+\frac{f^{(2)}f^2}{144 r^2}+\frac{f' f^{(2)} f^2}{144 r}+\frac{f^3}{36 r^4} \nonumber \\
& -\frac{Qf^3}{24 r^2}-\frac{f' f^3}{72 r^3}-\frac{f^{(2)} f^3}{144 r^2}-\frac{f^4}{72r^4}  \,, \\
\ma{T}^{(l)}_{10} = & \, \frac{Q}{2}-\frac{1}{6 r^2}+\frac{f'}{3 r}+\frac{f^{(2)}}{12}+\frac{f}{6 r^2}
\frac{Q^2}{16}-\frac{1}{120 r^4}-\frac{1}{48} Q' f'+\frac{Q f'}{24 r}+\frac{f'^2}{120r^2}+\frac{1}{48} Q f^{(2)} \nonumber \\
& -\frac{f' f^{(2)}}{120 r}+\frac{{f^{(2)}}^2}{480}-\frac{1}{240}f' f^{(3)} -\frac{1}{48} Q^{(2)} f-\frac{Q' f}{12 r}+\frac{f'f}{40 r^3}-\frac{f^{(2)} f}{30 r^2}-\frac{7 f^{(3)} f}{240 r} \nonumber \\
& -\frac{1}{240}f^{(4)} f+\frac{f^2}{120 r^4}  \,, \\
\ma{T}^{(l)}_{11} = & \, \frac{f}{60 r^4}-\frac{Q f}{24 r^2}-\frac{1}{48} Q' f' f-\frac{f'f}{24 r^3}+\frac{f'^2 f}{240 r^2}-\frac{1}{48} Q f^{(2)} f-\frac{f'f^{(2)} f}{40 r} -\frac{1}{240} {f^{(2)}}^2 f \nonumber \\
& -\frac{1}{480} f' f^{(3)}f+\frac{Q f^2}{24 r^2} +\frac{Q' f^2}{24 r} +\frac{f' f^2}{30r^3}+\frac{f^{(2)} f^2}{40 r^2}+\frac{f^{(3)} f^2}{80 r} +\frac{1}{480}f^{(4)} f^2-\frac{f^3}{60 r^4}\,.
}

\end{document}